\documentstyle[amssymb,prd,aps,epsfig]{revtex}

\newcommand{\be}{\begin{equation}}
\newcommand{\ee}{\end{equation}}
\newcommand{\ba}{\begin{eqnarray}}
\newcommand{\ea}{\end{eqnarray}}

\begin{document}
\draft
\title{{\bf Dimensional Reduction of the Abelian-Higgs Carroll-Field-Jackiw Model}}
\author{H. Belich$^{b,d}$, M.M. Ferreira Jr.$^{b,c}$ and J.A. Helay\"{e}l-Neto$^{a,b}
$\thanks{{\tt e-mails:} {\tt belich@cbpf.br, manojr07@aol.com,
helayel@cbpf.br.}}}
\address{$^{a}${\it Centro Brasileiro de Pesquisas F\'{i}sicas (CBPF)},\\
Coordena\c{c}\~{a}o de Teoria de Campos e Part\'{i}culas (CCP), \\
Rua Dr. Xavier Sigaud, 150 - Rio de Janeiro - RJ 22290-180 - Brazil.\\
$^{b}${\it Grupo de F\'{i}sica Te\'{o}rica Jos\'{e} Leite Lopes, }\\
Petr\'{o}polis - RJ - Brazil.\\
$^{c}${\it Universidade Federal do Maranh\~{a}o (UFMA)}, \\
Departamento de F\'{i}sica, Campus Universit\'{a}rio do Bacanga, S\~{a}o\\
Luiz - MA, 65085-580 - Brazil. \\
$^{d}$ {\it Universidade Federal do Esp\'{i}rito Santo (UFES)},\\
Departamento de F\'{i}sica e Qu\'{i}mica, Av. Fernando Ferrarim, S/N\\
Goiabeiras, Vit\'{o}ria - ES, 29060-900 - Brazil}
\maketitle

\begin{abstract}
Taking as a starting point a Lorentz non-invariant Abelian-Higgs model
defined in 1+3 dimensions, we carry out its dimensional reduction to $D=1+2$%
, obtaining a new planar model composed by a Maxwell-Chern-Simons-Proca
gauge sector, a massive scalar sector, and a mixing term (involving the
fixed background $v^{\mu }$) that imposes the Lorentz violation to the
reduced model. The propagators of the scalar and massive gauge field are
evaluated and the corresponding dispersion relations determined. Based on
the poles of the propagators, a causality and unitarity analysis is carried
out at tree-level. One then shows that the model is totally causal, stable
and unitary.
\end{abstract}


\section{ \ Introduction\-}

The point of view that some quantum field theories could be effective models
from more fundamental theories has been enhanced with the advent of
Supersymmetry and Supergravity, and more recently, with superstrings and
branes. In the end of \ 90%
\'{}%
s, some works \cite{Kostelec1} have demonstrated that a spontaneous
violation of Lorentz symmetry can take place in the context of string
theories. Some time later, the spontaneous violation of CTP\ and Lorentz
symmetries was adopted as a possibility to define some CPT and
Lorentz-violating models which can be taken as the low-energy limit of an
extension of the standard model defined at the Planck scale \cite{Colladay}.
This master model undergoes a spontaneous symmetry breaking, generating an
effective action that incorporates CPT and Lorentz violation and keeps
unaffected the $SU(3)\times SU(2)\times U(1)$ gauge structure of the
underlying theory. The Lorentz violation takes place at the level of
particle transformations, whereas at the level of observer rotations and
boots the effective model remains Lorentz invariant. Such a difference comes
from the role played by the CPT-violating background term, $v_{\mu },$ seen
as a four-vector under an observer Lorentz transformation and as a set of
four scalars in a particle frame. Moreover, the Lorentz covariance is
maintained as a feature of the underlying extended model, a consequence of
spontaneous character of the symmetry breaking. This fact is of relevance in
the sense it indicates that the effective model may preserve some properties
of the original theory, like causality and stability. Although Lorentz
symmetry is closely connected to stability and causality in modern field
theories, a model endowed with the latter properties in the absence of the
former should be in principle acceptable and meaningful on physical grounds.

Lorentz-violating theories have been in focus of recent and intensive
investigation. Such models have been presently adopted as an attempt to
explain the observation of ultra-high energy cosmic rays with energies
beyond the Greisen-Zatsepin-Kuzmin (GZK) cutoff $\left( E_{GZK}\simeq
4.10^{19}eV\right) $ \cite{Coleman}, \cite{Cosmic}, once such kind of
observation could be potentially taken as one evidence of Lorentz-violation.
The rich phenomenology of fundamental particles has also been considered as
a natural environment to the search for indications of breaking of these
symmetries \cite{Coleman},\cite{Particles}, indicating possible limitations
associated with such violation. Another point of interest refers to the
issue of space-time varying coupling constants \cite{Coupling}, which has
been reassessed in the light of Lorentz-violating theories, with interesting
connections with the construction of supergravity models. Moreover,
measurements of radio emission from distant galaxies and quasars put in
evidence that the polarizations vectors of the radiation emitted are not
randomly oriented as naturally expected. This peculiar phenomenon suggests
that the space-time intervening between the source and observer may be
exhibiting some sort of optical activity (birefringence), whose origin is
unknown \cite{Astrophys}.

The pure gauge sector of the Lorentz-violating low-energy effective model is
composed basically by two types of terms with respect to CTP-symmetry: (i)
the even CPT\ term, $k_{\alpha \beta \gamma \delta }F^{\alpha \beta
}F^{\gamma \delta },$\ where the coupling $k_{\alpha \beta \gamma \delta }$\
appears as a double traceless tensor with the same symmetries of the Riemann
tensor, and $F^{\gamma \delta }$ is the field strength; (ii) the odd CPT
term, $\epsilon _{\mu \nu \kappa \lambda }v^{\mu }A^{\nu }F^{\kappa \lambda
},$\ where $\epsilon _{\mu \nu \kappa \lambda }$\ is the 4-dimensional
Levi-Civita symbol and $v^{\mu }$\ is a fixed four-vector acting as a
background. This odd-CPT term (a Chern-Simons-like mass term) was first
considered in the context of a classical electrodynamics by
Carroll-Field-Jackiw \cite{Jackiw}, setting up a simple way to realize the
CPT- and Lorentz-breakings in the framework of the Maxwell theory. In spite
of predicting several interesting new properties and phenomenology, the
Carroll-Field-Jackiw (CFJ)\ model is plagued with some serious problems,
like absence of stability and causality in the case of a purely timelike
background, $v_{\mu }=($v$_{0},0).$ Even so, this theory has been object of
much attention in several different aspects, like the following ones: (i)
the birefringence (optical activity of the vacuum), induced by the fixed
background \cite{Jackiw},\cite{Astrophys}, (ii) the investigation of
radiative corrections \cite{Chung}, (iii) the consideration of spontaneous
breaking of U(1)-symmetry in this framework \cite{Belich1}, (iv) the search
for a supersymmetric Lorentz-violating extension model \cite{Belich2}, (v)
the study of vacuum Cerenkov radiation \cite{Lehnert}, photon decay process 
\cite{Adam2}, and the development of CFJ electrodynamics in a pre-metric
framework \cite{Itin}. $\ $

The quest for a Lorentz-violating model able to preserve the algebra of
supersymmetry (SUSY) was first addressed by Berger \& Kostelecky \cite
{Berger}. They have shown that a supersymmetric matter model in the presence
of a Lorentz-violating term could be achieved with success. Following a
different approach (starting from the degrees of freedom of the gauge
sector), the work of Ref. \cite{Belich2} has recently built up a
supersymmetric minimal extension of the Carroll-Field-Jackiw model,
obtaining also a non-polynomial extension compatible with $N=1$\ SUSY. On
other hand, the issue of the SSB was first addressed in Ref. \cite{Belich1},
where the spectrum was thoroughly discussed and electrically charged
vortices were found out.

Such a broad interest on the Carroll-Field-Jackiw model has triggered the
investigation of a similar model in a lower dimensional context. In this
way, the dimensional reduction (to 1+2 dimensions) of the
Carroll-Field-Jackiw model\ \cite{Jackiw} was successfully realized \cite
{Manojr1}, resulting in a planar theory composed of a Maxwell-Chern-Simons
gauge field $\left( A_{\mu }\right) $, a massless scalar field $\left(
\varphi \right) $, and a coupling term, $\varphi \epsilon _{\mu \nu \kappa
}v^{\mu }\partial ^{\nu }A^{\kappa }$, responsible for the Lorentz
violation. The reduced model has revealed to preserve causality, stability
and unitarity (in the gauge sector) both for a space- and time-like
backgrounds (without any restriction) \cite{Manojr1}, which bypasses the
lack of positivity and causality manifest in the 4-dimensional original
model. Such a result has put in evidence that this reduced model can undergo
a consistent quantization program (for both timelike and spacelike
backgrounds). Another interesting issue refers to the classical
electrodynamics concerning this planar Lagrangian, investigated initially at
the level of the motion equations taken at the static limit. Preliminary
results \cite{Manojr2} show that a purely timelike background induces the
behavior of a massless electrodynamics (in the electric sector), while a
pure spacelike background appears as a factor of strong anisotropy
promotion. The study of the scalar potential $\left( A_{0}\right) $\
solutions reveals the existence of a region where it is negative, which
favors the attainment of an electron-electron attractive potential, fact of
relevance in connection with condensed matter physics and recently confirmed
at least for a purely timelike background \cite{Manojr3}.

\ In this work, we aim at constructing and investigating a planar
Lorentz-violating model endowed with the Higgs sector. An extension of the
Carroll-Field-Jackiw model in $\left( 1+3\right) $\ dimensions, including a
scalar sector that yields spontaneous symmetry breaking (Higgs sector) \cite
{Belich1}, was recently developed and analyzed, providing an Abelian-Higgs
gauge model with violation of Lorentz symmetry. The planar counterpart of
this Abelian-Higgs model can be obtained by means of a dimensional reduction
(to 1+2 dimensions). The main\ motivation to study this kind of model is
twofold: (i) the relevance of considering a Lorentz-violating planar model
with spontaneous U(1)-symmetry breaking, which opens up the possibility of
analyzing the physical consistency of a Lorentz-violating theoretical
framework endowed with a Higgs sector in (1+2) dimensions; (ii) the need of
obtaining screened solutions, which is associated with condensed matter
systems, where one usually works with short range solutions. The presence of
the Higgs sector makes feasible promising investigations on vortex
configurations \cite{Belich3}, which may be of interest in connection with
anisotropic condensed matter systems.

\ In the present work, however, we really focus attention on the first
point: starting from the Abelian-Higgs model developed in Ref. \cite{Belich1}%
, we perform its dimensional reduction, having as outcome a planar Quantum
Electrodynamics (QED$_{3}$) described by a Maxwell-Chern-Simons gauge field, 
$A_{\mu },$\ by a massive Klein-Gordon field, $\varphi ,$\ and by the scalar
sector $\left( \phi \right) $\ minimally coupled to the gauge field, from
which the Higgs sector stems from. The $\varphi -$field also works out as
the coupling constant in the term that mixes the gauge field to the fixed
3-vector, $v^{\mu }$. A fourth-order scalar potential, $V,$\ then induces a
spontaneous symmetry breaking, which yields the appearance of the Higgs
scalar and a Proca mass component to the gauge field. Having established the
new planar Lagrangian, one then devotes some effort for the evaluation of
the propagators of the gauge and scalar fields, which requires the
definition of a closed algebra composed of eleven spin operators.\
Afterwards, the physical consistency of this model is investigated, with
causality, stability and unitarity being analyzed at the classical level.
Despite the presence of non-causal modes $\left( k^{2}<0\right) $\ coming
from the dispersion relations, the evaluation and analysis of the group and
front velocities is taken as a suitable criterium for assuring the
causality. Here, as it occurs in the reduced version \cite{Manojr1} of the
Maxwell-Carroll-Field-Jackiw model, the model reveals to be totally stable,
causal and unitary for both time- and space-like backgrounds, at the
classical level, bypassing the absence of stability and causality exhibited
by the original CFJ\ model. Once the unitarity is guaranteed, this model may
undergo a consistent quantization program, which is an important requirement
to the application of this model to describe physical systems.

This work is outlined as follows. In Sec II, we first perform the
dimensional reduction of the\ Abelian-Higgs Carroll-Field-Jackiw model,
obtaining the corresponding Lorentz-violating planar model. Afterwards, the
spontaneous symmetry breaking is considered and the propagators of the gauge
and scalar fields are evaluated. The knowledge of the propagators allows the
investigation of the\ physical consistency of this model. In Sec. III, the
stability and the causal structure of the model are analyzed, starting form
the dispersion relations extracted from the propagators. \ In Sec. IV, the
unitarity is suitably analyzed via the method of the residues (evaluated at
the poles of the propagators) of the current-current saturated propagator.
Finally, in Sec.V, we present our Concluding Remarks.

\section{The Dimensionally Reduced Model}

The starting point is a typical scalar electrodynamics, defined in $\left(
1+3\right) $ dimensions, endowed with the Carroll-Field-Jackiw term, as
written in Ref. \cite{Belich1}:

\begin{equation}
{\cal L}_{1+3}=-\frac{1}{4}F_{\hat{\mu}\hat{\nu}}F^{\hat{\mu}\hat{\nu}}+%
\frac{1}{4}\varepsilon ^{\hat{\mu}\hat{\nu}\hat{\kappa}\hat{\lambda}}v_{\hat{%
\mu}}A_{\hat{\nu}}F_{\hat{\kappa}\hat{\lambda}}+(D^{\hat{\mu}}\phi )^{\ast
}D_{\hat{\mu}}\phi -V(\phi ^{\ast }\phi )+A_{\hat{\nu}}J^{\hat{\nu}},
\label{action1}
\end{equation}
where the $\hat{\mu}$ runs from $0$ to $3,$ $D_{\hat{\mu}}=(\partial _{\hat{%
\mu}}+ieA_{\hat{\mu}})$ is the covariant derivative and $V(\phi ^{\ast }\phi
)=m^{2}\phi ^{\ast }\phi +\delta (\phi ^{\ast }\phi )^{2}$ represents the
scalar potential responsible for spontaneous symmetry breaking ($m^{2}<0$
and $\delta >0).$ This model is gauge invariant but does not preserve the
Lorentz and CTP symmetries.

In order to investigate this model in $\left( 1+2\right) $ dimensions, it is
necessary to perform its dimensional reduction, \ which consists effectively
in adopting the following ansatz over any 4-vector: (i) one keeps unaffected
the time and also the first two space components; (ii) one freezes the third
space dimension by splitting it from the body of the new 3-vector, ascribing
to it a scalar character; at the same time one requires that the new
quantities $\left( \chi \right) $, defined in $\left( 1+2\right) $
dimensions, do not depend on the third spacial dimension:\ $\partial
_{_{3}}\chi \longrightarrow 0.$ Applying this prescription\ to the gauge
4-vector, $A^{\hat{\mu}},$ and the fixed external 4-vector, $v^{\hat{\mu}},$
one \ has: 
\begin{eqnarray}
A^{\hat{\nu}} &\longrightarrow &(A^{\nu };\text{ }\varphi ),  \label{RP1} \\
v^{\hat{\mu}} &\longrightarrow &(v^{\mu };\text{ }s),  \label{RP2}
\end{eqnarray}
where: $A^{\left( 3\right) }=\varphi $, $v^{\left( 3\right) }=s$\ are two
scalars, and $\mu =0,1,2$. Carrying out this prescription for Eq. (\ref
{action1}), one then obtains: \ \ \ \ \ \ 
\begin{eqnarray}
{\cal L}_{1+2} &=&-\frac{1}{4}F_{\mu \nu }F^{\mu \nu }-\frac{s}{2}%
\varepsilon _{\mu \nu \kappa }A^{\mu }\partial ^{\nu }A^{\kappa }+\varphi
\varepsilon _{\mu \nu \kappa }v^{\hat{\mu}}\partial ^{\nu }A^{\kappa }+\frac{%
1}{2}\partial _{\mu }\varphi \partial ^{\mu }\varphi +(D^{\mu }\phi )^{\ast
}D_{\mu }\phi -e^{2}\varphi ^{2}\phi ^{\ast }\phi  \nonumber \\
&&-V(\phi ^{\ast }\phi )+A_{\nu }J^{\nu }+\varphi J.  \label{Lagrange2}
\end{eqnarray}
{\bf \ }The scalar, $\varphi ,$\ endowed with dynamics, is a typical
Klein-Gordon massless field, whereas $s$ is a constant\ scalar (without
dynamics), which acts as the Chern-Simons mass. The scalar field also
appears as the coupling constant that links the fixed $v^{\mu }$ to the
gauge sector of the model by means of the new term:\ $\varphi \varepsilon
_{\mu \nu k}v^{\mu }\partial ^{\nu }A^{k}.$ In spite of being covariant in
form, this kind of term breaks the Lorentz symmetry, since the 3-vector $%
v^{\mu }$ does not present dynamics. The presence of the Chern-Simons term
in Lagrangian (\ref{Lagrange2}), will also amount to the breakdown of the
parity and time reversal symmetries.

In adopting the dimensional reduction prescription as specified above ($%
\partial _{_{3}}\chi \longrightarrow 0$), we better clarify that the
integration over the $x_{3}$-coordinate, taken as usually to be compact,
will produce the length dimension that can be suitably absorbed into the
field and coupling constant redefinitions so as to yield the right canonical
dimensions for the fields and gauge coupling constant in (1+2)-D. This means
therefore that the Lagrangian{\bf \ }${\cal L}_{1+2}${\bf \ }naturally
carries the right canonical dimensions in 3D once its corresponding
4-dimensional master action has been fixed up. The dimensional reduction
procedure produces the right dimensional factor in such a way that the mass
dimensions turn out to be the ordinary ones.

Concerning the gauge invariance, it is noteworthy to state that the reduced
theory is gauge invariant under the reduction procedure (\ref{RP1}), (\ref
{RP2}). Indeed, the fact that all fields and the gauge parameter do not
depend on the third spatial coordinate $(x_{3})$\ guarantees that the scalar
field, $\varphi $, is a gauge invariant field in (1+2)D. On the other hand,
the scalar $s$, identified with $v^{\left( 3\right) }$, is constant mass
parameter; this shows that the term $\varepsilon _{\mu \nu k}A^{\mu
}\partial ^{\nu }A^{k}$\ is a genuine Chern-Simons term, gauge-invariant up
to a surface term. So, the reduction prescription here implemented allows
that the gauge symmetry of the action in (1+3)D survives in the planar
regime. Therefore, both the actions in four and three space-time dimensions
are gauge invariant modulo surface terms.

According to the prescription of dimensional reduction here adopted, a
comment is worthy: in the case the 4-dimensional background is purely
spacelike and orthogonal to the (1+2) dimensional subspace, that is $v^{\hat{%
\mu}}=(0,0,0,$v$),$\ there appears no sign of Lorentz-violation in the
reduced Lagrangian (\ref{Lagrange2}),\ once we are left with the genuine
Chern-Simons topological mass term.

We now proceed carrying out the spontaneous symmetry breaking, that takes
place when the scalar field exhibits a non null vacuum expectation value: $%
\langle \phi \phi \rangle =-m^{2}/2\delta $. Adopting the following
parametrization, $\phi =(\varkappa +\eta /\sqrt{2})e^{i\rho \eta /\sqrt{2}}$%
, we obtain (for $\rho =0):$

\begin{eqnarray}
{\cal L}_{1+2}^{Broken} &=&-\frac{1}{4}F_{\mu \nu }F^{\mu \nu }+\frac{s}{2}%
\varepsilon _{\mu \nu \kappa }A^{\mu }\partial ^{\nu }A^{\kappa }-\varphi
\varepsilon _{\mu \nu \kappa }v^{\mu }\partial ^{\nu }A^{\kappa }+\frac{1}{2}%
\partial _{\mu }\varphi \partial ^{\mu }\varphi -e^{2}\varkappa ^{2}\varphi
^{2}+e^{2}\varkappa ^{2}A_{\mu }A^{\mu }  \nonumber \\
&&+\frac{1}{2}\partial _{\mu }\eta \partial ^{\mu }\eta +\frac{2}{\sqrt{2}}%
e^{2}\varkappa \eta A_{\mu }A^{\mu }+\frac{e^{2}}{2}\eta ^{2}A_{\mu }A^{\mu
}+m^{2}(\varkappa +\eta /\sqrt{2})^{2}+\delta (\varkappa +\eta /\sqrt{2}%
)^{4}.
\end{eqnarray}
Retaining only tree-level terms, we obtain an action in an explicitly
quadratic form, 
\begin{equation}
\Sigma _{1+2}=\int d^{3}x\frac{1}{2}\biggl\{A^{\mu }[Z_{\mu \nu }]A^{\nu
}-\varphi (\square +M_{A}^{2})\varphi -\varphi \left[ \epsilon _{\mu \alpha
\nu }v^{\mu }\partial ^{\alpha }\right] A^{\nu }+A^{\mu }\left[ \epsilon
_{\nu \alpha \mu }v^{\nu }\partial ^{\alpha }\right] \varphi \biggr\},
\label{action3}
\end{equation}
where the mass of the scalar field is the same as the Proca mass ($%
M_{A}^{2}=2e^{2}\varkappa ^{2}).$ Here, the mass dimension of the physical
parameters and tensors are:\ $\left[ A^{\mu }\right] =\left[ \varphi \right]
=1/2,$ $\left[ v^{\mu }\right] =\left[ s\right] =1,\left[ T_{\mu }\right] =%
\left[ Z_{\mu \nu }\right] =2.$ The action (\ref{action3}) can also be read
in a matrix form:\ 

\begin{equation}
\Sigma _{1+2}=\int d^{3}x\frac{1}{2}\left( 
\begin{array}{cc}
A^{\mu } & \varphi
\end{array}
\right) \left[ 
\begin{array}{cc}
Z_{\mu \nu } & T_{\mu } \\ 
-T_{\nu } & -(\square +M_{A}^{2})
\end{array}
\right] \ \left( 
\begin{array}{c}
A^{\nu } \\ 
\varphi
\end{array}
\right) .  \label{actionQ}
\end{equation}
Now, we define the operators we shall be dealing with:\ 
\begin{eqnarray}
Z_{\mu \nu } &=&\square \theta _{\mu \nu }+s{\rm \ }S_{\mu \nu
}+M_{A}^{2}g_{\mu \nu },\text{ \ \ \ }T_{\mu }=S_{\nu \mu }v^{\nu },
\label{Pre1} \\
\ S_{\mu \nu } &=&\varepsilon _{\mu \kappa \nu }\partial ^{\kappa },\text{ \
\ }\theta _{\mu \nu }=\eta _{\mu \nu }-\omega _{\mu \nu },\text{ }\ \ \ 
\text{\ \ }\omega _{\mu \nu }=\frac{\partial _{\mu }\partial _{\nu }}{%
\square },  \label{Pre2}
\end{eqnarray}
where $\theta _{\mu \nu }$ and $\omega _{\mu \nu }$ are respectively the
dimensionless transverse and longitudinal projectors.

The propagators of the gauge and scalar fields are given by the inverse of
the square matrix, $Q,$\ associated with the action (\ref{actionQ}). The
propagator matrix, $\Delta ,$is then written as: 
\begin{equation}
\Delta =Q^{-1}=\frac{-1}{(\square +M_{A}^{2})Z_{\mu \nu }+T_{\mu }T_{\nu }}%
\left[ 
\begin{array}{cc}
-(\square +M_{A}^{2}) & T_{\nu } \\ 
-T_{\mu } & Z_{\mu \nu }
\end{array}
\right] ,
\end{equation}
whose components are given by: \ 
\begin{eqnarray}
\left( \Delta _{11}\right) ^{\mu \nu } &=&(\square +M_{A}^{2})\left[ -Z_{\mu
\nu }(\square +M_{A}^{2})+T_{\mu }T_{\nu }\right] ^{-1},  \label{Prop1} \\
\left( \Delta _{22}\right) &=&-Z_{\mu \nu }\left[ Z_{\mu \nu }(\square
+M_{A}^{2})-T_{\mu }T_{\nu }\right] ^{-1}, \\
\left( \Delta _{12}\right) ^{\mu } &=&-T_{\nu }\left[ Z_{\mu \nu }(\square
+M_{A}^{2})-T_{\mu }T_{\nu }\right] ^{-1}, \\
\left( \Delta _{21}\right) ^{\nu } &=&T_{\mu }\left[ Z_{\mu \nu }(\square
+M_{A}^{2})-T_{\mu }T_{\nu }\right] ^{-1}.  \label{Prop4}
\end{eqnarray}
The terms $\Delta _{11}$, $\Delta _{22}$\ correspond to the propagators of
the gauge and scalar fields, while the terms $\Delta _{12},$\ $\Delta _{21}$%
\ are the mixed propagators $\langle \varphi A_{\mu }\rangle ,$\ $\langle
A_{\mu }\varphi \rangle $, which describe a scalar mediator turning into a
gauge mediator and vice-versa. In order to explicitly obtain these
propagators, it is necessary to invert the matrix components individually.
For this purpose, one needs to\ create some new operators, in such a way a
closed operator algebra can be defined. In this sense, we define the
following tensor operators: 
\begin{equation}
Q_{\mu \nu }=v_{\mu }T_{\nu },\text{ \ }\Lambda _{\mu \nu }=v_{\mu }v_{\nu },%
\text{ \ \ }\Sigma _{\mu \nu }=v_{\mu }\partial _{\nu },\text{ \ }\Phi _{\mu
\nu }=T_{\mu }\partial _{\nu },
\end{equation}
which fulfill some useful relations: 
\begin{eqnarray}
S_{\mu \nu }T_{\text{ \ }}^{\nu }T^{\alpha } &=&\square v_{\mu }T^{\alpha
}-\lambda T^{\alpha }\partial _{\mu }=\square Q_{\mu }^{\text{ \ }\alpha
}-\lambda \Phi _{\text{ \ }\mu }^{\alpha }, \\
Q_{\mu \nu }Q^{\alpha \nu } &=&T^{2}v^{\alpha }v_{\mu }=T^{2}\Lambda _{\text{
\ }\mu }^{\alpha },\text{ }Q_{\mu \nu }\Phi ^{\nu \alpha }=T^{2}v_{\mu
}\partial ^{\alpha }=T^{2}\Sigma _{\mu }^{\text{ \ }\alpha }, \\
\lambda &\equiv &\Sigma _{\mu }^{\;\mu }=v_{\mu }\partial ^{\mu }\;,\;\text{%
\ }T^{2}=T_{\alpha }T^{\alpha }=(v^{2}\square -\lambda ^{2}).
\end{eqnarray}
Their mass dimensions are: $\left[ \Lambda _{\mu \nu }\right] =2,$ $\left[
Q_{\mu \nu }\right] =3,$ $\left[ \Sigma _{\mu \nu }\right] =2\;,\;\left[
\Phi _{\mu \nu }\right] =3$.

The inversion of $\Delta _{11}$ is realized following the traditional
prescription, $\left( \Delta _{11}^{-1}\right) _{\mu \nu }\left( \Delta
_{11}\right) ^{\nu \alpha }=\delta _{\mu }^{\alpha },$ where the operator $%
\left( \Delta _{11}\right) ^{\nu \alpha }$ is the most general tensor
operator composed of 2-rank combinations of the one-forms $T_{\mu },v_{\mu
},\partial _{\alpha }.$ In this sense, the operators $Q_{\mu \nu },Q_{\nu
\mu },$\ $\Sigma _{\mu \nu },\Sigma _{\nu \mu },\Phi _{\mu \nu },\Phi _{\nu
\mu }$\ must all be considered,\ leading to a linear combination of eleven
terms: 
\begin{eqnarray}
\left( \Delta _{11}\right) ^{\nu \alpha } &=&a_{1}\theta ^{\nu \alpha
}+a_{2}\omega ^{\nu \alpha }+a_{3}S^{\nu \alpha }+a_{4}\Lambda ^{\nu \alpha
}+a_{5}T^{\nu }T^{\alpha }+a_{6}Q^{\nu \alpha }+a_{7}Q^{\alpha \nu
}+a_{8}\Sigma ^{\nu \alpha }+a_{9}Q^{\alpha \nu }  \nonumber \\
&&+a_{10}\Phi ^{\nu \alpha }+a_{11}\Phi ^{\alpha \nu }.
\end{eqnarray}
The closure of the operator algebra involving these operators is contained
in Table I of Ref. \cite{Manojr1}, whose application leads to the following
propagator of the gauge field:

\begin{eqnarray}
\left( \Delta _{11}\right) ^{\mu \nu }\text{ } &=&\frac{(\square +M_{A}^{2})%
}{\boxplus }\theta ^{\mu \nu }+\frac{(\square +M_{A}^{2})\boxtimes \boxplus
-\lambda ^{2}s^{2}M_{A}^{2}\square }{M_{A}^{2}(\square +M_{A}^{2})\boxtimes
\boxplus }\omega ^{\mu \nu }-\frac{s}{\boxplus }S^{\mu \nu }-\frac{%
s^{2}\square ^{2}}{(\square +M_{A}^{2})\boxtimes \boxplus }\Lambda ^{\mu \nu
}  \nonumber \\
&&+\frac{(\square +M_{A}^{2})}{\boxtimes \boxplus }T^{\mu }T^{\nu }-\frac{%
s\square }{\boxtimes \boxplus }\left[ Q^{\mu \nu }-Q^{\nu \mu }\right] +%
\frac{\lambda s^{2}\square }{(\square +M_{A}^{2})\boxtimes \boxplus }\left[
\Sigma ^{\mu \nu }+\Sigma ^{\nu \mu }\right] -\frac{s\lambda }{\boxtimes
\boxplus }\left[ \Phi ^{\mu \nu }-\Phi ^{\nu \mu }\right] ,
\end{eqnarray}
where: $\boxtimes =\left[ (\square +M_{A}^{2})^{2}-T^{2}+s^{2}\square \right]
,\boxplus =(\square +M_{A}^{2})^{2}+s^{2}\square .$

According to Eqs. (\ref{Prop1}-\ref{Prop4}), the propagators $\left( \Delta
_{12}\right) ^{\alpha }$\ and $\left( \Delta _{21}\right) ^{\alpha }\ $can
be written in terms of the $\Delta _{11}$-gauge propagator,

\begin{equation}
\left( \Delta _{12}\right) ^{\alpha }=-\frac{T_{\mu }}{(\square +M_{A}^{2})}%
\left( \Delta _{11}\right) ^{\mu \alpha },\text{ \ }\left( \Delta
_{21}\right) ^{\alpha }=\frac{T_{\mu }}{(\square +M_{A}^{2})\boxtimes }%
\left( \Delta _{11}\right) ^{\alpha \mu },
\end{equation}
which leads to the following propagator expressions:

\begin{equation}
\left( \Delta _{12}\right) ^{\alpha }=\frac{-1}{(\square
+M_{A}^{2})\boxtimes }\left[ (\square +M_{A}^{2})T^{\alpha }+s\square
v^{\alpha }-s\lambda \partial ^{\alpha }\right] ,
\end{equation}

\begin{equation}
\left( \Delta _{21}\right) ^{\alpha }=\frac{1}{(\square +M_{A}^{2})\boxtimes 
}\left[ (\square +M_{A}^{2})T^{\alpha }-s\square v^{\alpha }+s\lambda
\partial ^{\alpha }\right] .
\end{equation}
As for the scalar field propagator, it can be put in the tensor form below: 
\begin{equation}
\left( \Delta _{22}\right) =-\left[ (\square +M_{A}^{2})-T_{\mu }(Z_{\mu \nu
})^{-1}T_{\nu }\right] ^{-1},
\end{equation}
which can be easily solved by taking the inverse of the tensor $Z_{\mu \nu
}, $ 
\begin{equation}
\left( Z_{\mu \nu }\right) ^{-1}=\frac{(\square +M_{A}^{2})}{\boxplus }%
\theta ^{\mu \nu }-\frac{s}{\boxplus }S^{\mu \nu }+\frac{1}{M_{A}^{2}}\omega
^{\mu \nu }.
\end{equation}
Making use of the following outcome, $T_{\mu }(Z^{-1})^{\mu \nu }T_{\nu
}=(\square +M_{A}^{2})T^{2}/\boxplus $, a simple scalar propagator arises:

\[
\left( \Delta _{22}\right) =-\frac{\boxplus }{\boxtimes (\square +M_{A}^{2})}%
. 
\]
Now, we can write the propagators here obtained in momentum-space. The
photon propagator takes on its final form:

\begin{eqnarray}
\langle A^{\mu }\left( k\right) A^{\nu }\left( k\right) \rangle \text{ } &=&i%
\biggl\{-\frac{(k^{2}-M_{A}^{2})}{\boxplus (k)}\theta ^{\mu \nu }+\frac{%
(k^{2}-M_{A}^{2})\boxtimes \boxplus -\lambda ^{2}s^{2}M_{A}^{2}k^{2}}{%
M_{A}^{2}(k^{2}-M_{A}^{2})\boxtimes (k)\boxplus (k)}\omega ^{\mu \nu }-\frac{%
s}{\boxplus }S^{\mu \nu }+\frac{s^{2}k^{4}}{(k^{2}-M_{A}^{2})\boxtimes
(k)\boxplus (k)}\Lambda ^{\mu \nu }  \nonumber \\
&&-\frac{(k^{2}-M_{A}^{2})}{\boxtimes (k)\boxplus (k)}T^{\mu }T^{\nu }+\frac{%
sk^{2}}{\boxtimes (k)\boxplus (k)}\left[ Q^{\mu \nu }-Q^{\nu \mu }\right] +%
\frac{i(v.k)s^{2}k^{2}}{(k^{2}-M_{A}^{2})\boxtimes (k)\boxplus (k)}\left[
\Sigma ^{\mu \nu }+\Sigma ^{\nu \mu }\right]  \nonumber \\
&&-\frac{is(v.k)}{\boxtimes (k)\boxplus (k)}\left[ \Phi ^{\mu \nu }-\Phi
^{\nu \mu }\right] \biggr\},  \label{gauge1}
\end{eqnarray}
while the scalar and the mixed propagators read as 
\begin{equation}
\text{ }\langle \varphi \varphi \rangle \text{ }=i\frac{\boxplus (k)}{%
\boxtimes (k)(k^{2}-M_{A}^{2})},  \label{Prop_phi}
\end{equation}
\begin{equation}
\langle A^{\alpha }\varphi \rangle \text{ }=\frac{-i}{(k^{2}-M_{A}^{2})%
\boxtimes (k)}\left[ (k^{2}-M_{A}^{2})T^{\alpha }+sk^{2}v^{\alpha }+s(v\cdot
k)k^{\alpha }\right] ,
\end{equation}
\begin{equation}
\langle \varphi A^{\alpha }\rangle =\frac{i}{(k^{2}-M_{A}^{2})\boxtimes (k)}%
\left[ (k^{2}-M_{A}^{2})T^{\alpha }-sk^{2}v^{\alpha }+s(v\cdot k)k^{\alpha }%
\right] ,
\end{equation}
where: $\boxtimes (k)=k^{4}-\left( 2M_{A}^{2}+s^{2}-v\cdot v\right)
k^{2}+M_{A}^{4}-\left( v\cdot k\right) ^{2},$ $\boxplus
(k)=(k^{2}-M_{A}^{2})^{2}-s^{2}k^{2}.$

Since we are committed to the calculation of physical quantities such as the
mass spectrum and the residues of the propagators at their poles, we take
the viewpoint of working in the unitary gauge. Local U(1)-symmetry has been
spontaneously broken, so that we could have also chosen to adopt the R$_{\xi
}$-type gauge, for which the would-be-Goldstone scalar propagates (its pole
is however gauge-dependent) and the longitudinal part of the gauge-field
propagator displays the same gauge-dependent pole. However, this gauge is
more convenient for the study of more formal aspects, like
renormalizability, for example. To get information on the mass spectrum and
on the physical character of the propagator poles, the choice of the unitary
gauge seems to be more natural. It is the gauge symmetry, even though
spontaneously broken, that allows us to adopt either choice, once at the
level of the S-matrix the results will be perfectly equivalent.

\section{Causality and Stability Analysis}

Despite Lorentz symmetry to be a cornerstone in field theory,
Lorentz-violating theoretical models may be acceptable once there occurs
preservation of two physical essential properties: causality and stability
(energy positivity). The poles of the propagators can be taken as a suitable
starting point to get information about causality, stability and unitarity
of the correlated model. The causality analysis, at tree-level, is related
to the sign of the propagator poles, given in terms of $k^{2},$ in such a
way that one must have $k^{2}\geq 0$ in order to preserve the causality
(preventing the existence of tachyons). The families of poles at $k^{2}$
coming from the propagators expressions are given below: 
\begin{equation}
k^{2}=M_{A}^{2};\text{ }\boxplus (k)=0\text{\ };\text{ \ }\boxtimes (k)=0;
\label{K2-poles}
\end{equation}
from which we extract the dispersion relations associated with each one. In
the case of \thinspace $k^{2}=M_{A}^{2}$, we obtain a very simple dispersion
relation, $k_{0}^{2}=M_{A}^{2}+{\bf k}^{2},$ which obviously establishes
both a causal and stable mode.

Concerning the equation $\boxplus (k)=0$, we attain background-independent
roots:

\begin{equation}
\ k_{\pm }^{2}=M_{A}^{2}+\frac{s^{2}}{2}\pm \frac{|s|}{2}\sqrt{%
s^{2}+4M_{A}^{2}}.  \label{polo4}
\end{equation}
The causality is preserved at these poles, since we have: $k_{\pm }^{2}>0.$
The stability of these modes is also assured.

As for the poles of $\boxtimes (k)=0,$ we obtain: 
\begin{equation}
\ k_{\pm }^{2}=M_{A}^{2}+\frac{s^{2}}{2}-\frac{v\cdot v}{2}\pm \frac{1}{2}%
\sqrt{(s^{2}-v\cdot v)(s^{2}-v\cdot v+4M_{A}^{2})+4(v\cdot k)^{2}}.
\label{polo5}
\end{equation}
In the case of a purely time-like background, $v^{\mu }=($v$_{0},{\bf 0}),$
these poles assume the following form: 
\begin{equation}
k_{\pm }^{2}=M_{A}^{2}+s^{2}/2\pm \sqrt{s^{4}/4+M_{A}^{2}s^{2}+\text{v}%
_{0}^{2}{\bf k}^{2}},  \label{polo6a}
\end{equation}
from which we note that the pole $k_{+}^{2}$ is always causal and stable
whereas the pole $k_{-}^{2}$,\ beyond to be non-causal ($k_{-}^{2}$ $<0),$
seems to be non stable. Hence, the first analysis of relevance refers to the
stability (positivity of the energy) of the mode{\bf \ }$k_{-}^{2}.${\bf \ }%
A simple investigation reveals that the expression for the energy, $%
k_{0-}^{2}=M_{A}^{2}+s^{2}/2+{\bf k}^{2}\pm \sqrt{s^{4}/4+M_{A}^{2}s^{2}+%
\text{v}_{0}^{2}{\bf k}^{2}},$\ is always positive for any value of{\bf \ }$%
{\bf k}^{2}$ whenever the single condition $s^{2}>$v$_{0}^{2}$\ is
fulfilled. Once the stability is assured, it turns out feasible to show that
the non-causal character of \ this\ last pole ($k_{-}^{2}$\ $<0)$\ is not
decisive to spoil the causality of the model. In order to do it, one takes
as essential point the evaluation of the group and the front velocities
associated with the pole $k_{-}^{2}.$\ Adopting $k^{\mu }=(k_{0},0,k_{2}),$\
the group velocity $(v_{g}=dk_{0-}/dk_{2})$\ results equal to 
\begin{equation}
v_{g}=\frac{k_{2}}{k_{0-}}\frac{\sqrt{s^{4}/4+M_{A}^{2}s^{2}+\text{v}%
_{0}^{2}k_{2}^{2}}-\text{v}_{0}^{2}/2}{\sqrt{s^{4}/4+M_{A}^{2}s^{2}+\text{v}%
_{0}^{2}k_{2}^{2}}}.
\end{equation}
Such a velocity is always less than 1, once the energy expression for $%
k_{0-} $\ does not possess any pole (it is positive definite for any value
of k$^{2} $). In the limit $k_{2}\rightarrow \infty ,$ one has $v_{g}=1.$
From the phase velocity $\left( v_{ph}=k_{0-}/k_{2}\right) ,$ one can obtain
the front velocity $\left( v_{f}=\lim_{\text{k}\rightarrow \infty
}|v_{ph}|\right) $, which stands for a sensitive factor for signal
propagation \cite{Adam}, \cite{Brillouin}. Considering Eq. (\ref{polo6a}),
one easily notes that it yields a unitary front velocity $\left(
v_{f}=1\right) $ in the limit k$_{2}\rightarrow \infty ,$ which regarded
jointly with $v_{g}\leq 1$ constitutes a suitable criterium to assure
causality at classical level.

For a purely space-like background, $v^{\mu }=(0,{\bf v}),$ Eq. (\ref{polo5}%
) reads as 
\begin{equation}
\ k_{\pm }^{2}=M_{A}^{2}+s^{2}/2+{\bf v}^{2}/2\pm \frac{1}{2}\sqrt{(s^{2}+%
{\bf v}^{2})(s^{2}+{\bf v}^{2}+4M_{A}^{2})+4({\bf v}.{\bf k})^{2}}.
\label{polo6b}
\end{equation}
In this case, we have the same behavior as in the purely time-like
situation, that is, the pole $k_{+}^{2}$ is always causal and stable,
whereas the pole $k_{-}^{2}$ \ is non-causal ($k_{-}^{2}<0).${\bf \ }Now,
one can show that the stability of this mode can be assured $\left(
k_{0-}^{2}>0\right) $\ without any restriction over the parameters. Adopting 
$k^{\mu }=(k_{0},0,k_{2}),$\ the group velocity $(v_{g}=dk_{0-}/dk_{2})$\ is
then given as follows: 
\begin{equation}
v_{g}=\frac{k_{2}}{k_{0-}}\frac{\sqrt{(s^{2}+{\bf v}^{2})(s^{2}+{\bf v}%
^{2}+4M_{A}^{2})+4{\bf v}_{2}^{2}k_{2}^{2}}-\text{{\bf v}}_{2}^{2}}{\sqrt{%
(s^{2}+{\bf v}^{2})(s^{2}+{\bf v}^{2}+4M_{A}^{2})+4{\bf v}_{2}^{2}k_{2}^{2}}}%
.
\end{equation}
This expression implies that $v_{g}<1$\ for any value of k$_{2}$ and $%
v_{g}=1 $ in the limit $k_{2}\rightarrow \infty $. Analogously, it may be
shown that the front velocity is unitary $\left( v_{f}=1\right) $, a
sufficient condition to prevent the spectrum of the model from the presence
of non-causal modes and to assure the causality of physical signals.
Therefore, despite the presence of non-causal poles $\left(
k_{-}^{2}<0\right) $ in both time- and space-like cases, the conditions $%
v_{g}<1$ and $v_{f}=1$ exclude the appearance of tachyons.

\section{Unitarity}

The unitarity analysis of the reduced model at tree-level is here carried
out through the saturation of the propagators with external currents, which
must be implemented both to the scalar $(J)$ and gauge $(J^{\mu })$\
currents, once the model presents these two sectors. In such a way, we write
individually the two saturated propagators ($SP$) at the following form:

\begin{eqnarray}
SP_{\langle A_{\mu }A_{\nu }\rangle } &=&J^{\ast \mu }\langle A_{\mu
}(k)A_{\nu }(k)\rangle \text{ }J^{\nu },  \label{Sat1} \\
SP_{\langle \varphi \varphi \rangle } &=&J^{\ast }\langle \varphi \varphi
\rangle J.  \label{Sat2}
\end{eqnarray}
While the gauge current $(J^{\mu })$\ satisfies the conservation law $\left(
\partial _{\mu }J^{\mu }=0\right) ,$\ the scalar current $(J)$\ does not
fulfill any constraint. Into the context of this method, the unitarity
analysis is assured whenever the imaginary part of the residues of $SP$\ at
the poles of each propagator is positive.

\subsection{Scalar Sector}

The unitarity analysis of the scalar sector is performed by means of Eq.(\ref
{Sat2}), or more explicitly: 
\begin{equation}
SP_{\langle \varphi \varphi \rangle }=J^{\ast }\frac{i\boxplus (k)}{%
\boxtimes (k)(k^{2}-M_{A}^{2})}J.  \label{Sat3}
\end{equation}
This expression presents three poles: $M_{A}^{2}$, and $k_{+}^{2},k_{-}^{2}$
(the roots of $\boxtimes (k)=0).$ At the purely time-like case, $v^{\mu }=($v%
$_{0},{\bf 0}),$ the poles $k_{\pm }^{2}$ are exactly the ones given by Eq. (%
\ref{polo6a}). The residues of $SP_{\langle \varphi \varphi \rangle }$,
evaluated at these three poles, are positive-definite, in such a way the
unitarity of the scalar sector, in the time-like case, is completely assured.

At the purely space-like case, $v^{\mu }=(0,{\bf v}),$ the poles of Eq. (\ref
{Sat3}) are $M_{A}^{2}$ and the ones given by Eq. (\ref{polo6b}). The
residues of $SP_{\langle \varphi \varphi \rangle },$\ carried out at these
three poles, provide a positive-definite imaginary part, so that the
unitarity at the space-like case, is generically preserved. So, we conclude
that the unitarity of the scalar sector is ensured without any restrictions.

\subsection{Gauge Field}

As for the gauge field, the continuity equation, $k_{\mu }J^{\mu }=0,$
reduces to six the number of terms of the photon propagator that contributes
to the evaluation of the saturated propagator:\ 
\begin{eqnarray}
SP &=&J_{\mu }^{\ast }(k)\biggl\{\frac{i}{D}[(\square
+M_{A}^{2})^{2}\boxtimes g^{\mu \nu }-s(\square +M_{A}^{2})\boxtimes S^{\mu
\nu }-s^{2}\square ^{2}\Lambda ^{\mu \nu }+(\square +M_{A}^{2})^{2}T^{\mu
}T^{\nu }  \nonumber \\
&&-s\square (\square +M_{A}^{2})(Q^{\mu \nu }-Q^{\nu \mu })]\biggr\}J_{\nu
}(k),
\end{eqnarray}
where: $D=(\square +M_{A}^{2})\boxtimes \boxplus .$\ In this case, the
current components exhibit the form $J^{\mu }=(j^{0},0,\frac{k_{0}}{k_{2}}%
j^{(0)})$\ whenever one adopts as momentum $k^{\mu }=(k_{0},0,k_{2}).$
Writing this expression in momentum-space, one obtains: 
\begin{equation}
SP=J^{\ast \mu }(k)\biggl\{B_{\mu \nu }\biggr\}J^{\nu }(k),
\end{equation}
where: $D=-(k^{2}-M_{A}^{2})\boxtimes (k)\boxplus (k),$ $\boxtimes
(k)=k^{4}-\left( 2M_{A}^{2}+s^{2}-v\cdot v\right) k^{2}+M_{A}^{4}-\left(
v\cdot k\right) ^{2},$ $\boxplus (k)=(k^{2}-M_{A}^{2})^{2}-s^{2}k^{2}.$

\subsubsection{Timelike case:}

For a purely timelike 3-vector, $v^{\mu }=($v$_{0},{\bf 0}),$ $k^{\mu
}=(k_{0},0,k_{2}),$ the tensor $B_{\mu \nu }$\ is given as follows:{\bf \ } 
\begin{equation}
B_{\mu \nu }(k)=\frac{i}{D\left( k\right) }\left[ 
\begin{array}{ccc}
C^{4}\boxtimes -s^{2}\text{v}_{0}^{2}k^{4} & -isk^{(2)}C^{2}[\boxtimes -%
\text{v}_{0}^{2}k^{2}] & isC^{2}k^{\left( 1\right) }(\boxtimes -\text{v}%
_{0}^{2}k^{2}) \\ 
isk^{(2)}C^{2}[\boxtimes -\text{v}_{0}^{2}k^{2}] & -C^{4}[\boxtimes +\text{v}%
_{0}^{2}k_{2}^{2}] & -C^{2}[is\boxtimes k_{0}-C^{2}\text{v}%
_{0}^{2}k^{(1)}k^{(2)}] \\ 
-isC^{2}k^{(1)}(\boxtimes -\text{v}_{0}^{2}k^{2}) & C^{2}[is\boxtimes
k_{0}-C^{2}\text{v}_{0}^{2}k^{(1)}k^{(2)}] & -C^{4}[\boxtimes +\text{v}%
_{0}^{2}k_{1}^{2}]
\end{array}
\right] ,
\end{equation}
where it was used the short notation: $C^{2}=(k^{2}-M_{A}^{2}).$

We start by performing unitarity analysis for the first pole, \ $%
k^{2}=M_{A}^{2},$ for which the residue of the matrix $B_{\mu \nu }$ can be
reduced to a very simple form: 
\begin{equation}
B_{\mu \nu }(M_{A}^{2})=i\frac{M_{A}^{2}\text{v}_{0}^{2}}{%
s^{2}[s^{2}M_{A}^{2}+\text{v}_{0}^{2}k_{2}^{2}]}\left[ 
\begin{array}{ccc}
1 & 0 & 0 \\ 
0 & 0 & 0 \\ 
0 & 0 & 0
\end{array}
\right] ,
\end{equation}
which implies a positive saturation $\left( SP>0\right) ,$ and preservation
of unitarity.

For the poles of $\boxtimes (k)=0,$ given by Eq. (\ref{polo6a}), we obtain
the following residue matrix:

\begin{equation}
B_{\mu \nu }(k_{\pm }^{2})=iR_{\pm }\text{v}_{0}^{2}\left[ 
\begin{array}{ccc}
s^{2}k_{\pm }^{4} & -isk_{\pm }^{2}(k_{\pm }^{2}-M_{A}^{2})k^{(2)} & 0 \\ 
isk_{\pm }^{2}(k_{\pm }^{2}-M_{A}^{2})k^{(2)} & (k_{\pm
}^{2}-M_{A}^{2})^{2}k_{2}^{2} & 0 \\ 
0 & 0 & 0
\end{array}
\right] ,
\end{equation}
where $R_{\pm }$ is the residue of $1/D\left( k\right) $ evaluated at $%
k_{\pm }^{2},$ namely: 
\[
R_{\pm }=2\text{v}_{0}^{2}{\bf k}^{2}\left( s^{2}/2\pm \sqrt{%
s^{4}/4+4M_{A}^{2}s^{2}+4\text{v}_{0}^{2}{\bf k}^{2}}\right) \left( \pm 
\sqrt{s^{4}/4+4M_{A}^{2}s^{2}+4\text{v}_{0}^{2}{\bf k}^{2}}\right) , 
\]
which implies $(R_{\pm })>0$. The eigenvalues of the matrix above are: $%
\lambda _{1}=0;\lambda _{2}=0;\lambda _{3}=s^{2}k_{\pm
}^{4}+k_{2}^{2}(k_{\pm }^{4}-2M_{A}^{2}k_{\pm }^{2}+M_{A}^{4}).$ Since $%
\lambda _{3}$ is a positive eigenvalue, the saturation results positive $%
\left( SP>0\right) ,$ and the unitarity is assured.

For the poles of $\boxplus (k)=0,$ given by Eq. (\ref{polo4}), we obtain the
following residue matrix:

\begin{equation}
B_{\mu \nu }|_{(k^{2}=k_{\pm }^{2})}=iR_{\pm }\left[ 
\begin{array}{ccc}
C^{4}\boxtimes -s^{2}\text{v}_{0}^{2}k_{\pm }^{4} & -isk^{(2)}C^{2}[%
\boxtimes -\text{v}_{0}^{2}k_{\pm }^{2}] & 0 \\ 
isk^{(2)}C^{2}[\boxtimes -\text{v}_{0}^{2}k_{\pm }^{2}] & 0 & 
-isC^{2}\boxtimes k_{0} \\ 
0 & isC^{2}\boxtimes k_{0} & -C^{4}\boxtimes
\end{array}
\right] ,
\end{equation}
where: $C^{2}=(k_{\pm }^{2}-M_{A}^{2}),\boxtimes (k_{\pm }^{2})=-$v$%
_{0}^{2}k_{2}^{2}$, and $R_{\pm }$ is the residue of $1/D\left( k\right) $
evaluated at $k_{\pm }^{2},$ so that $R_{\pm }>0$. This matrix leads to a
null saturation\ $\left( SP=0\right) $\ whenever saturated with the external
current $J^{\mu }=(j^{0},0,\frac{k_{0}}{k_{2}}j^{(0)}),$\ which implies
preservation of unitarity. The trivial saturation at these poles shows that
the modes given by Eqs. (\ref{polo4}) are non-dynamical for the pure
time-like background; therefore, they do not stand for a physical excitation.

\subsubsection{Spacelike Case:}

For a pure spacelike fixed vector, $v^{\mu }=(0,0,V),$ $k^{\mu
}=(k_{0},0,k_{2}),$ the 2-rank tensor $B_{\mu \nu }$ can be put in the
following matrix form: 
\begin{equation}
B_{\mu \nu }(k)=\frac{i}{D\left( k\right) }\left[ 
\begin{array}{ccc}
C^{4}(\boxtimes -V^{2}k_{1}^{2}) & -iC^{2}[s\boxtimes
k^{(2)}+iV^{2}k_{0}k^{(1)}] & isC^{2}(\boxtimes +V^{2}k^{2})k^{\left(
1\right) } \\ 
iC^{2}[s\boxtimes k^{(2)}-iV^{2}k_{0}k^{(1)}] & -C^{4}[\boxtimes
+V^{2}k_{0}^{2}] & -isC^{2}[\boxtimes +V^{2}k^{2}]k_{0} \\ 
-isC^{2}(\boxtimes +V^{2}k^{2})k^{\left( 1\right) } & isC^{2}[\boxtimes
+V^{2}k^{2}]k_{0} & -C^{4}\boxtimes -s^{2}V^{2}k^{4}
\end{array}
\right] .
\end{equation}
First, we perform the unitarity analysis at the pole, \ $k^{2}=M_{A}^{2},$
for which the residue of the matrix $B_{\mu \nu }$ can be simplified to a
simple form: 
\begin{equation}
B_{\mu \nu }(M_{A}^{2})=i\frac{s^{2}V^{2}M_{A}^{4}}{%
s^{2}[s^{2}M_{A}^{2}+V^{2}M_{A}^{2}+V^{2}k_{2}^{2}]}\left[ 
\begin{array}{ccc}
0 & 0 & 0 \\ 
0 & 0 & 0 \\ 
0 & 0 & 1
\end{array}
\right] ,
\end{equation}
which clearly implies a positive saturation ($SP>0)$ and preservation of
unitarity.

For the poles of $\boxtimes (k)=0,$ given by Eq. (\ref{polo6b}), we obtain
the following residue matrix:

\begin{equation}
B_{\mu \nu }(k_{\pm }^{2})=-iR_{\pm }V^{2}\left[ 
\begin{array}{ccc}
0 & 0 & 0 \\ 
0 & (k_{\pm }^{2}-M_{A}^{2})k_{0}^{2} & is(k_{\pm }^{2}-M_{A}^{2})k_{\pm
}^{2}k_{0} \\ 
0 & -is(k_{\pm }^{2}-M_{A}^{2})k_{\pm }^{2}k_{0} & s^{2}k_{\pm }^{4}
\end{array}
\right] ,
\end{equation}
where $R_{\pm }$ is the residue of $1/D\left( k\right) $ evaluated at $%
k_{\pm }^{2},$ so that $R_{\pm }<0$. The eigenvalues of the matrix above
are: $\lambda _{1}=0;\lambda _{2}=0;\lambda _{3}=s^{2}k_{\pm
}^{4}+k_{0}^{2}(k_{\pm }^{4}-2M_{A}^{2}k_{\pm }^{2}+M_{A}^{4}).$ Since $%
\lambda _{3}$ is a positive eigenvalue, the saturation results positive $%
\left( SP>0\right) ,$ and the unitarity is assured.

For the poles of $\boxplus (k)=0,$ given by Eq. (\ref{polo4}), we obtain the
following residue matrix:

\begin{equation}
B_{\mu \nu }|_{(k^{2}=k_{\pm }^{2})}=iR_{\pm }\left[ 
\begin{array}{ccc}
C^{4}\boxtimes & -isC^{2}\boxtimes k^{(2)} & 0 \\ 
isC^{2}\boxtimes k^{(2)} & 0 & isC^{2}(\boxtimes +V^{2}k_{\pm }^{2})k_{0} \\ 
0 & -isC^{2}(\boxtimes +V^{2}k_{\pm }^{2})k_{0} & -C^{4}\boxtimes
-s^{2}V^{2}k_{\pm }^{4}
\end{array}
\right] ,
\end{equation}
where: $C^{2}=(k_{\pm }^{2}-M_{A}^{2}),\boxtimes (k_{\pm
}^{2})=-V^{2}k_{0}^{2}$, and $R_{\pm }$ is the residue of $1/D\left(
k\right) $ evaluated at $k_{\pm }^{2},$ so that $R_{\pm }>0$. This matrix,
whenever saturated with the external current $J^{\mu }=(j^{0},0,\frac{k_{0}}{%
k_{2}}j^{(0)}),$ leads to a trivial saturation\ $\left( SP=0\right) $, which
is compatible with unitarity requirements. The vanishing of $SP$ at these
poles indicates that the modes given by Eq. (\ref{polo4}) are non-dynamical
for the pure space-like background too.

This is the whole lot of our investigations as long as causality and
unitarity at tree-level are concerned. We finish remarking that the reduced
model preserves unitarity, for both space- and time-like backgrounds,
without any restriction.

\bigskip

\section{Concluding Remarks}

We have carried out the dimensional reduction to (1+2 dimensions) of an
Abelian-Higgs gauge model with the Carroll-Field-Jackiw Lorentz-violating
term (defined in 1+3 dimensions). One attains a planar model composed of a
Maxwell-Chern-Simons-Proca gauge sector, a massive scalar sector and a
mixing term that couples the gauge field to the fixed background. The
propagators of this model are evaluated and the causality, stability and
unitarity are analyzed. Concerning stability, it is entirely ensured
whenever the auxiliary condition $s^{2}>v_{0}^{2}$\ is valid. Furthermore,
we have shown that the overall model preserves causality and unitarity for
both timelike and spacelike backgrounds, the same outcome attained in Ref. 
\cite{Manojr1}. This result encourages us to push forward our idea of
applying the $(1+2)-D$ counterpart of the $(1+3)$ Lorentz-broken models to
discuss issues related to physical planar systems, once this model can be
submitted to a consistent quantization scheme. Though fermions play a
central role if we are committed to applications to low-dimensional
Condensed Matter systems, we have not introduced them in our presentation.
The reason is that the Lorentz breaking and its immediate consequence for
the causality and unitarity are classically felt only by the charged scalars
and gauge fields. The introduction of the fermions is the next natural step
and it remains to be worked out the influence (at the planar level) of the
background vector, $v^{\mu },$ in yielding Lorentz-breaking terms in the
fermionic sector. This matter is now under consideration.

A natural extension of the present work is the investigation of its
classical equations of motion (for potentials and field strengths) and their
corresponding solutions, in a similar way as it appears in Ref. \cite
{Manojr2}. Thus, the structure of the resulting electrodynamics associated
with the planar Lagrangian (\ref{Lagrange2}) can be readily determined, at
least in the static regime. Preliminary calculations reveal that the
solutions for the field strengths and potentials have a very similar
structure to the ones of the pure MCS-Proca electrodynamics. This issue is
now under development \cite{ClassicalHiggs}. A study of vortex
configurations (for time- and space-like backgrounds) was also carried out
simultaneously to the analysis of the classical aspects alluded to here,
revealing that this model is also endowed with stable vortex configurations 
\cite{Belich3}.

Another point to be investigated concerns the evaluation of the
electron-electron interaction in the context of this planar model. This
matter may be analyzed in much the same way adopted in Ref. \cite{Manojr3},
where one has evaluated the $e^{-}e^{-}$\ interaction potential for the case
of the Lorentz-violating MCS electrodynamics of Ref. \cite{Manojr1}. It was
then verified that the interaction potential may be attractive for some
parameter values and exhibits a logarithmic potential near and far from the
origin. In the case of the\ Lorentz-violating MCS-Proca electrodynamics here
developed, one expects the maintenance of the attractive character at the
same time the resulting electron-electron potential is supposed \ to be
totally screened due to presence of the additional Proca mass parameter.

\end{document}